\documentstyle[aps,multicol,epsf]{revtex}
\begin{document}
\title{Extremal Segments in Random Sequences}
\author{Yacov Kantor{\dag} and Deniz Erta\c s{\ddag}}
\address{{\dag}School of Physics and Astronomy, Tel Aviv University,
Tel Aviv 69 978, Israel\\
{\ddag}Department of Physics, Massachusetts Institute of
Technology, Cambridge, MA 02139, U.S.A.}
\date{\today}
\maketitle
\begin{abstract}
We investigate the probability for the largest segment
in with total displacement $Q$ in an $N$-step random
walk to have length $L$. Using analytical, exact enumeration,
and Monte Carlo methods, we reveal the complex structure of
the probability distribution in the large $N$ limit.
In particular, the size of the longest loop has a distribution
with a square-root singularity at $\ell\equiv L/N=1$, an
essential singularity
at $\ell=0$, and a discontinuous derivative at $\ell=1/2$.
\end{abstract}
\begin{multicols}{2}
\narrowtext
Investigation of the ground states of randomly charged
polymers\cite{KK} suggests that in order to take maximal
advantage of condensation energy and to diminish the effects
of long range repulsion of the excess charges,
the polymer will select a necklace-like configuration,
consisting of a few large, almost neutral globules, connected
by narrow chains. In general this presents a complicated
energy minimization problem. Some aspects of the solution
can be determined by asking a simpler question:
What is the length of the {\it longest} segment of the
random sequence (RS) of charges that has total charge $Q$?
Alternatively, one can think of a one dimensional random
walk (RW) in which the longest segment with an end-to-end
distance $Q$ is to be found. The problem resembles certain
classical RW problems\cite{rchandra}, such as the problem
of first and last arrival to a given point, or the special
case of the last return to the starting point of the RW.
However, the search for the longest segment of the RW,
among all possible starting points, creates a more
complicated problem. We combine Monte Carlo (MC) and exact
enumeration studies, with some exact analytical results in
certain simple limits, to demonstrate some remarkable
properties of the distribution of the maximal--length
segments.

\begin{figure}
\epsfxsize=3.1in
\centerline{\epsffile{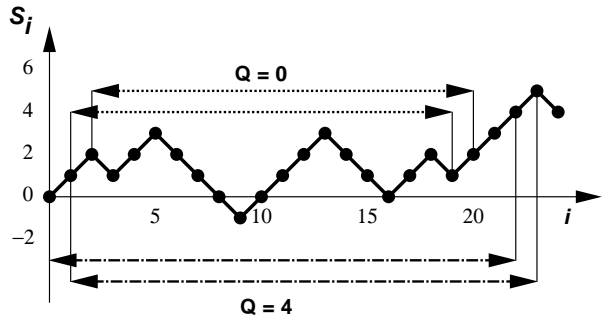}}
\medskip
\caption{Example of a RS. In this case, the longest 0--segments
have lengths $L=18$ (dotted lines), while the longest
4--segments (dot--dashed lines) have lengths $L=22$.}
\label{FigA}
\end{figure}

A RS is described by a sequence of random charges
$\{q_i\}$ ($i=1,\dots,N)$, where $q_i=\pm1$ with
equal probability. Fig.~\ref{FigA} depicts an example of the
accumulated charge $S_i=\sum_{j=1}^iq_j$ for a RS.
($S_0\equiv0$.) Every segment of the sequence between, say,
steps $i$ and $j$, has a certain charge $Q=S_j-S_i$. Such a
segment will be denoted as a $Q$--segment. (A 0--segment
corresponds to a loop in a RW, i.e., a segment for which
the positions at the beginning and the end coincide.)
Consider the set of all $Q$--segments for a fixed value of
$Q$. Our task is to find segments of largest length $L$
among these. Fig.~\ref{FigA} shows the longest 0--segments
and the longest 4--segments, in a RS with $N=24$. Clearly,
the longest $Q$--segment does not have to be unique. We
are interested in the probability $P_N(L,Q)$ that $L$ is
the length of the longest $Q$--segment in a RS of length
$N$. For $|Q|>0$, the set of $Q$--segments for a given
sequence may be empty, and therefore
$\sum_{L=0}^N P_N(L,|Q|>0)<1$.

\begin{figure}
\hbox to \hsize{\hfil
\epsfysize=3in\epsffile{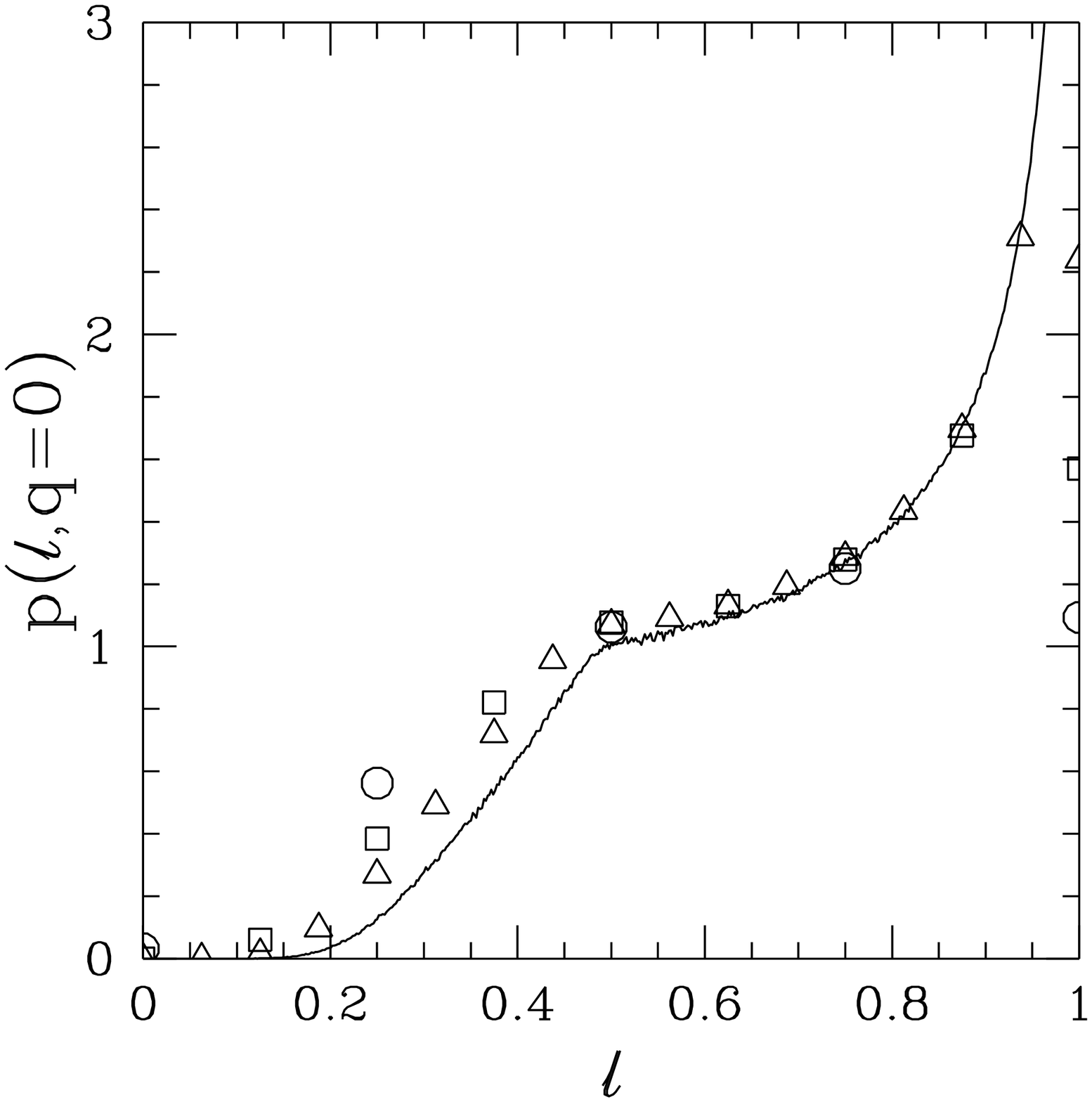}
\hfil\hfil
\epsfxsize=3in\epsffile{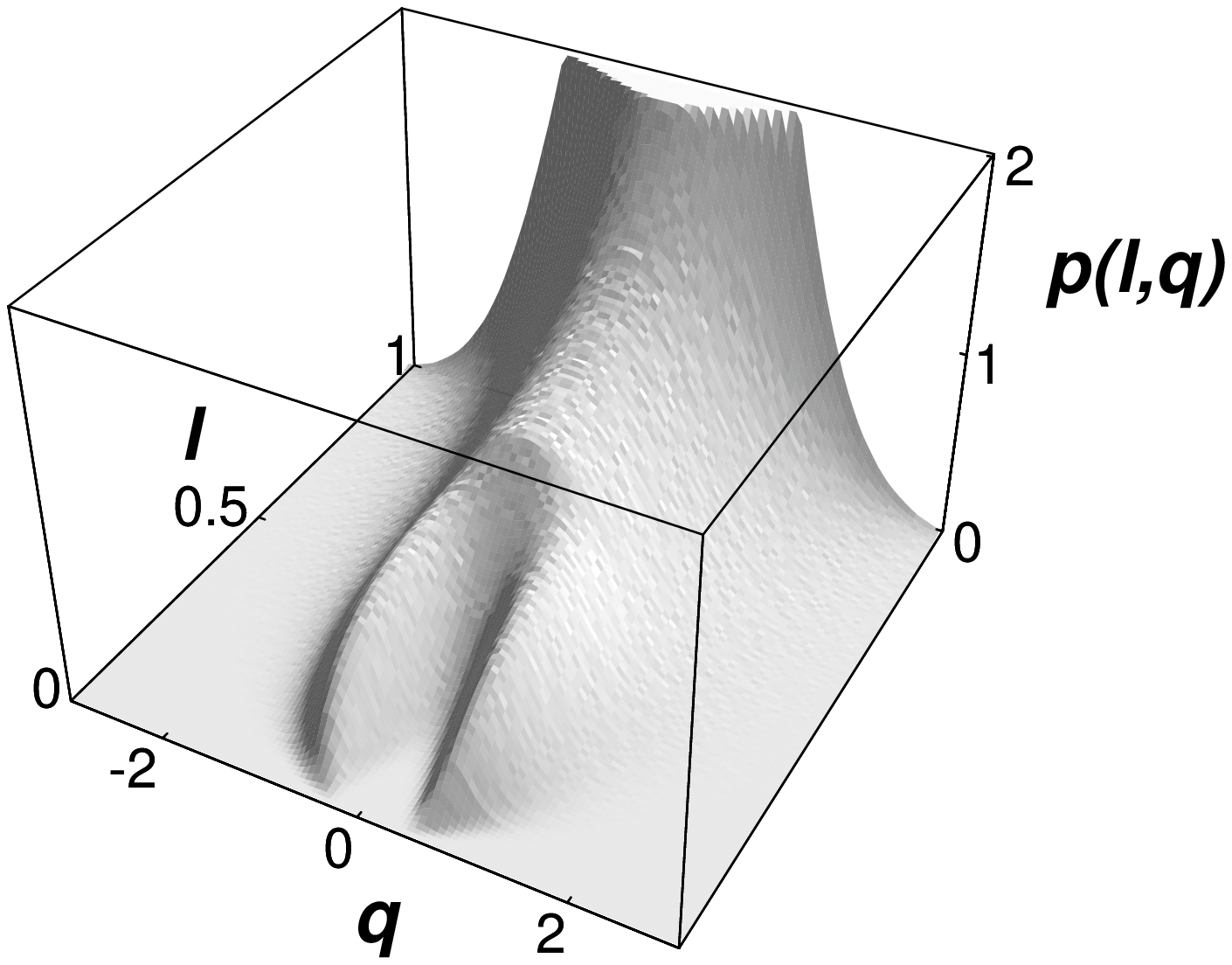}
\hfil}
\medskip
\caption{(a) Probability density of 0--segments as a function of
reduced length $\ell$. Circles, squares, and triangles depict the
exact enumeration results for $N=8$, 16, and 32, respectively.
The solid line shows results of MC simulations (see text).
(b) Probability density of $Q$--segments
as a function of reduced charge $q$ and reduced length $\ell$.
The results have been obtained from MC simulations (see text).}
\label{FigB}
\end{figure}

Most properties of RSs have simple continuum limits. For
example, the probability that an $N$-element RS has total
charge $Q$ (for even $N+Q$) is
\begin{eqnarray}
W_N(Q) &\quad=\quad& 2^{-N}\frac{N!}{[(N-Q)/2]![(N+Q)/2]!} \nonumber \\
 & {\displaystyle \mathop{=}_{N\to\infty}} & \;\;
\sqrt{\frac{2}{\pi N}}\exp(-Q^2/2N).
\end{eqnarray}
Similarly, we expect $P_N(L,Q)$ to approach a simple form
when $N,L,Q\to\infty$, while the reduced length
$\ell\equiv L/N$ and the reduced charge $q\equiv Q/\sqrt{N}$
are kept constant. In that (continuum) limit it is more
convenient to work with the {\it probability density}
$p(\ell,q)={N\over 2}\left[P_N(L,Q)+P_N(L+1,Q)\right]$.
(At most one of the two probabilities is nonzero, since
$P_N=0$ for odd $L+Q$.) In certain cases, $P_N$ can be
calculated exactly, especially for very small values of $L$
and $N-L$, and for arbitrary $Q$. For example,
$P_N(L=N,Q)=W_N(Q)$, while
$P_N(L=N-2,Q)={1\over 4}\left[W_{N-4}(Q+2)+2W_{N-4}(Q)
+W_{N-4}(Q-2)\right]$. Similarly, one can find expressions
for very small $L$\cite{tobepub}. However, we were unable
to find a general expression for arbitrary $N$, $L$ and $Q$.
We performed exact enumeration studies of $P_N(L,0)$ for
$N\le 36$. Results for few values of $N$ are shown in
Fig.~\ref{FigB}a. The results converge extremely fast
to the continuum distribution $p(\ell,0)$.
The solid curve in the same figure depicts
the results of a MC evaluation of the probability density
from $10^8$ randomly selected sequences of length $N=1000$.
\end{multicols}

\widetext

\begin{multicols}{2}

The probability density $p(\ell,0)$ shown in Fig.~\ref{FigB}a
has several remarkable properties: (a) MC results show
that $p$ at $\ell={1\over2}$ is very close to unity
($1.004\pm0.006$). At that point the slope of the
curve changes by an order of magnitude.
(b) For $\ell\to0$, the function exhibits an essential
singularity of the form $\sim \ell^{-2}\exp(-B/\ell)$,
where $B\approx1.7$. (c) For  $\ell\to1$, the
function diverges as $(1-\ell)^{-1/2}$.
Qualitatively, this behaviour can be understood as follows:
The length of the longest 0--segment strongly depends
on the overall charge $Q_o$ of the chain. For simplicity,
let us assume that it depends only on $Q_o$.
Then, for small values of $Q_o$ we can relate  $\ell=1-aQ_o^2/N$,
where $a$ is of order unity. For very large $Q_o$ and, thus,
for very small $\ell$, the length of the longest 0--segment
will be of order of a scale at which the random excursion of
the RW becomes comparable to the drift produced by $Q_o$,
i.e., when $L^{1/2}\approx L Q_o/N$, and therefore
$\ell\approx N/Q_o^2$. By applying the relation
$p(\ell,0)=(N/2)W_N(Q_o)|dQ_o/d\ell|$ in both limits, we correctly
reproduce the square-root divergence for $\ell\to1$, and
the $\exp({\rm const}/\ell)$ singularity for $\ell\to0$. (The
leading pre-exponential power is not reproduced correctly
in the latter case. A more involved argument\cite{tobepub}
also reproduces this power correctly.) It is interesting to note
that, by matching the asymptotic form of $p(\ell,0)$ near $\ell=1$
with $P_N(L,0)$ for $L=N-2$, we reproduce almost
the exact value of the prefactor, i.e., the discrete
distribution approaches its asymptotic (continuum) form
within a few steps of the extreme $L=N$.

Fig.~\ref{FigB}b depicts the full probability density  $p(\ell,q)$,
obtained from a MC evaluation of $10^7$ sequences of length $N=1024$.
This figure demonstrates further peculiarities of $p(\ell,q)$:
For fixed $\ell$, the $q$-dependence of $p$ is qualitatively
different for $\ell >{1\over2}$ and $\ell < {1\over 2}$. In the former
case, the distribution has a single peak at $q=0$, and the areas
$A_\ell\equiv\int_{-\infty}^{+\infty}dq\,p(\ell,q)$ under fixed-$\ell$
sections have the form ${\rm const}/\sqrt{1-\ell}$. In the latter case,
however, we see two peaks, and $A_\ell$ is approximately linear in
$\ell$ for $0.15 < \ell < 0.5$.

An interesting and potentially useful integral relation exists
between the probabilities $P_N(L,Q)$. The number of sequences in
which the longest $Q$--segment has length $L$ is
$2^NP_N(L,Q)$. In the particular case of $L\ge N/2$,
we can construct all such sequences as follows:
First we construct all sequences of length
$2(N-L)$  with longest $Q'$--segment of length $N-L$,
i.e., exactly half of its total length. There are
$2^{2(N-L)}P_{2(N-L)}(N-L,Q')$ such sequences.
Next, we construct all sequences of length $2L-N$ and
total charge $Q-Q'$. There are $2^{2L-N}W_{2L-N}(Q-Q')$
such sequences. Finally, by inserting any chain from the
second group into any chain from the first group at its
midpoint, and repeating this process for all possible
values of $Q'$, we will reproduce all the sequences of
total length $N$, with longest $Q$--segments of length
$L$. It is easy to verify that this process produces
every desired configuration once and only once. The
necessary condition, however, is that $L\ge N/2$, i.e.,
the longest $Q$--segment must include the midpoint of the
sequence. In the continuum limit, this relation
can be expressed as
\begin{eqnarray}\label{eintrel}
p(\ell,q)&=&{1\over\sqrt{4\pi(2\ell-1)(1-\ell)}}\nonumber \\
&&\;\times\int\limits_{-\infty}^{+\infty}\!\!dq'\,
{\rm e}^{-{\left(q-q'\sqrt{2(1-\ell)}\right)^2\over 2(2\ell-1)}}
p\left({1\over2},q'\right)\ ,
\end{eqnarray}
where the reduced variable $q'=Q'/\sqrt{2(N-L)}$, and the Gaussian
term in the integrand represents the continuum limit of
$W_{2L-N}(Q-Q')$. Equation (\ref{eintrel}) gives the probabilities
for any $\ell\ge{1\over2}$ in terms of their value at
$\ell={1\over2}$, and reduces to identity in the $\ell\to{1\over2}$
limit. By integrating both sides of Eq.(\ref{eintrel}) over $q$,
we find a relation between the areas $A_\ell$, for $\ell>{1\over 2}$:
\begin{equation}
A_\ell={1\over\sqrt{2(1-\ell)}}
\int\limits_{-\infty}^{+\infty}\!\!dq\,p\left({1\over2},q\right)\ ,
\end{equation}
which confirms the observation from the MC data that for
$\ell>{1\over2}$, $A_\ell$ simply increases as $1/\sqrt{1-\ell}$.
In the $\ell\to1$ limit, the variable
$q'$ disappears from the exponent in Eq.(\ref{eintrel}),
and the relation reduces to
\begin{equation}
p(\ell\to1,q)={A_{1/2}\over\sqrt{4\pi(1-\ell)}}{\rm e}^{-q^2/2}\ .
\end{equation}
This relation both confirms our contention that $p(\ell,0)$
has a square-root divergence $A/\sqrt{\pi(1-\ell)}$ with
$A\equiv{1\over 2}A_{1/2}$, and demonstrates that the
fixed--$\ell$ sections of the surface in Fig.\ \ref{FigB}b
approach a pure Gaussian shape when $\ell\to1$. In addition
to the MC study, we performed an exact enumeration study to
determine $A$ for sequences with $N\le 30$, and found that
it extrapolates to the value $1.011\pm0.001$, in perfect
agreement with the MC result: Definitely larger than unity,
but surprisingly close to it.

We did not find analogous integral relations for
$\ell<{1\over2}$. Here, the situation is complicated by the
fact that, in a given sequence, there may be several longest
$Q$-segments that are disjoint. The $q$--dependence of
$p(\ell,q)$ for small values of $\ell$ has a minimum at $q=0$.
The minimum disappears as $\ell$ increases, at
$\ell\approx{1\over2}$. Further analysis is necessary to
understand the behaviour of $p(\ell,q)$ in this region.

In conclusion, we demonstrated that the probability density
$p(\ell,q)$ has some peculiar and unexpected properties
and very rich behaviour, despite the apparent simplicity of
its formulation, and its similarity to classical RW problems.
More analysis is needed to fully understand various properties of
the extremal segments in a RS.

We thank M.~Kardar for helpful discussions.
This work was supported by the US--Israel BSF grant
No. 92--00026, and by the NSF through grant Nos.
DMR--87--19217 (at MIT's CMSE) and DMR 91--15491 (at Harvard).

\end{multicols}
\end{document}